\documentclass[twoside,twocolumn,9pt]{article}
\usepackage{extsizes}
\usepackage[super,sort&compress,comma]{natbib} 
\usepackage[version=3]{mhchem}
\usepackage[left=1.5cm, right=1.5cm, top=1.785cm, bottom=2.0cm]{geometry}
\usepackage{balance}
\usepackage{mathptmx}
\usepackage{sectsty}
\usepackage{graphicx} 
\usepackage{lastpage}
\usepackage[format=plain,justification=justified,singlelinecheck=false,font={stretch=1.125,small,sf},labelfont=bf,labelsep=space]{caption}
\usepackage{float}
\usepackage{fancyhdr}
\usepackage{fnpos}
\usepackage[english]{babel}
\addto{\captionsenglish}{%
  
}
\usepackage{array}
\usepackage{droidsans}
\usepackage{charter}
\usepackage[T1]{fontenc}
\usepackage[usenames,dvipsnames]{xcolor}
\usepackage{setspace}
\usepackage[compact]{titlesec}
\usepackage{hyperref}

\usepackage{epstopdf}

\definecolor{cream}{RGB}{222,217,201}

\begin{document}

\pagestyle{fancy}
\thispagestyle{plain}
\fancypagestyle{plain}{
\renewcommand{\headrulewidth}{0pt}
}

\makeFNbottom
\makeatletter
\renewcommand\LARGE{\@setfontsize\LARGE{15pt}{17}}
\renewcommand\Large{\@setfontsize\Large{12pt}{14}}
\renewcommand\large{\@setfontsize\large{10pt}{12}}
\renewcommand\footnotesize{\@setfontsize\footnotesize{7pt}{10}}
\makeatother

\renewcommand{\thefootnote}{\fnsymbol{footnote}}
\renewcommand\footnoterule{\vspace*{1pt}%
\color{cream}\hrule width 3.5in height 0.4pt \color{black}\vspace*{5pt}} 
\setcounter{secnumdepth}{5}

\makeatletter 
\renewcommand\@biblabel[1]{#1}            
\renewcommand\@makefntext[1]%
{\noindent\makebox[0pt][r]{\@thefnmark\,}#1}
\makeatother 
\renewcommand{\figurename}{\small{Fig.}~}
\sectionfont{\sffamily\Large}
\subsectionfont{\normalsize}
\subsubsectionfont{\bf}
\setstretch{1.125} 
\setlength{\skip\footins}{0.8cm}
\setlength{\footnotesep}{0.25cm}
\setlength{\jot}{10pt}
\titlespacing*{\section}{0pt}{4pt}{4pt}
\titlespacing*{\subsection}{0pt}{15pt}{1pt}

\fancyfoot{}
\fancyfoot[RO]{\footnotesize{\sffamily{1--\pageref{LastPage} ~\textbar  \hspace{2pt}\thepage}}}
\fancyfoot[LE]{\footnotesize{\sffamily{\thepage~\textbar\hspace{3.45cm} 1--\pageref{LastPage}}}}
\fancyhead{}
\renewcommand{\headrulewidth}{0pt} 
\renewcommand{\footrulewidth}{0pt}
\setlength{\arrayrulewidth}{1pt}
\setlength{\columnsep}{6.5mm}
\setlength\bibsep{1pt}

\makeatletter 
\newlength{\figrulesep} 
\setlength{\figrulesep}{0.5\textfloatsep} 

\newcommand{\topfigrule}{\vspace*{-1pt}%
\noindent{\color{cream}\rule[-\figrulesep]{\columnwidth}{1.5pt}} }

\newcommand{\botfigrule}{\vspace*{-2pt}%
\noindent{\color{cream}\rule[\figrulesep]{\columnwidth}{1.5pt}} }

\newcommand{\dblfigrule}{\vspace*{-1pt}%
\noindent{\color{cream}\rule[-\figrulesep]{\textwidth}{1.5pt}} }

\makeatother

\twocolumn[
  \begin{@twocolumnfalse}
\vspace{1em}
\sffamily
\begin{tabular}{m{0cm} p{17.5cm} }

& \noindent \LARGE{\textbf{The structure of disintegrating defect clusters in smectic C freely suspended films}} \\
\vspace{0.3cm} & \vspace{0.3cm} \\

 & \noindent \large{Ralf Stannarius$^{\dag}$\textit{$^{a}$}},
              \large{Kirsten Harth$^{\ast}$\textit{$^{b}$}}  \\ \\
\vspace{0.3cm} & \vspace{0.3cm} \\

& \noindent \normalsize{Disclinations or disclination clusters in smectic
C freely suspended films with topological charges larger than one are unstable. They disintegrate, preferably in a spatially symmetric fashion, into single defects with individual charges +1, which is the smallest positive topological charge allowed in polar vector fields. While the opposite process of defect annihilation is well-defined by the initial defect positions, a disintegration starts from a singular state and the following scenario including the emerging regular defect patterns must be selected by specific mechanisms.
We analyze experimental data and compare them with a simple model where the defect clusters adiabatically pass quasi-equilibrium solutions in one-constant approximation. It is found that the defects arrange in geometrical patterns that correspond very closely to superimposed singular defect solutions, without additional director distortions. The patterns expand by affine transformations where all distances between individual defects scale with the same time-dependent scaling factor proportional to the square-root of time.} \\

\end{tabular}

 \end{@twocolumnfalse} \vspace{0.6cm}
  ]

\renewcommand*\rmdefault{bch}\normalfont\upshape
\rmfamily
\section*{}
\vspace{-1cm}


\footnotetext{\textit{$^{a}$~Institute of Physics, Otto von Guericke University, D-39106 Magdeburg, Universit\"ats\-platz 2, Germany.}}
\footnotetext{\textit{$^{b}$~Department of Engineering, Brandenburg University of Applied Sciences, D-14770 Brandenburg an der Havel, Magdeburger Stra\ss{}e 50, Germany.}}

\footnotetext{$\dag$~ralf.stannarius@ovgu.de, $\ast$~kirsten.harth@th-brandenburg.de}




\section{Introduction}

Topological defects play a significant role in a huge variety of physical systems, including for example anisotropic soft matter \cite{Brinkman1982,Cladis1987,Poulin1998,Musevic2006,Tkalec2011,Musevic2019,Dolganov2022,Gan2022,Loudet2022}, biological matter \cite{Doostmohammadi2018,Ardaseva2022,Copenhagen2021,Meacock2021,Amiri2022}, superfluidic liquids \cite{Ruutu1996,Bauerle1996,Bauerle1996b}, quantum systems \cite{Weiler2008,Polkovnikov2011,He2016},  or even cosmology \cite{Volovik1977,Chuang1991a}. The close resemblance of topological defects in biological systems and in active nematics has been emphasized \cite{Fardin2021,Meacock2021,Doostmohammadi2018}. Coarsening of defect patterns is a common feature in the initial evolution of dynamic systems after symmetry-breaking phase transitions. The elementary processes of such coarsening scenarios are mutual annihilations of topological defect pairs with opposite charges, a scenario that has been considered in a variety of studies in different fields. It shall be mentioned that in liquid-crystalline systems in general, disclinations are quite easily prepared and observed in experiments,
and the annihilation of point defects in particular has been studied extensively. A recent review of this field can be found in Ref.~\cite{Harth2020}.

In a previous experiment with smectic C freely suspended films, the annihilation of oppositely charged defects was studied experimentally, and the role of the anisotropy of elastic constants and of macroscopic material flow in the films was demonstrated \cite{Missaoui2020}. The elastic anisotropy mainly manifests itself in a preferential pinning of the director near the defect cores.
A model developed earlier for nematic systems \cite{Vromans2016,Tang2017}
was employed and extended to describe the experimentally observed effects of mutual defect orientation on the annihilation dynamics. One key ingredient of this model is the fixed orientation of defects with topological strength $S=+1$ in such films, mentioned above as a consequence of the elastic anisotropy.

The motivation to study this particular system and geometry is that the smectic freely
suspended liquid-crystalline films offer clear advantages for systematic studies of defect interactions: First, the films allow to investigate defect dynamics in a quasi two-dimensional (2D) geometry. The air above and below the films does not affect local orientations and has only little influence on flow in the film plane. In addition, the observation of defect motions and orientations is straightforward by means of polarizing microscopy. The timescales of defect dynamics in the range of seconds are convenient for video recording, and the continuum equations governing the film dynamics are well developed. Several, but not all involved material parameters are known from other experiments.
Simple methods for the experimental preparation of defect patterns in such films have been established \cite{Pargellis1992,Muzny1992,Muzny1994,Wachs2014,Stannarius2016,Missaoui2020}.

The present study does not deal with the annihilation dynamics but focuses on the opposite scenario, the decomposition of high-strength
point defects. The basis of the following theoretical analysis is an earlier experimental investigation of smectic C freely suspended films with mutually repelling topological defects of equal charges \cite{Stannarius2016}.
The emerging single defects arrange in certain characteristic, highly symmetric patterns. The structure of these patterns has no obvious relation to the initial state. We derive a model for the pattern selection and explain why some typical patterns are found experimentally while others are avoided.

\section{Experimental observations}

In the smectic C phase, the mesogens are on average tilted with respect to the smectic layer plane. The molecular layers in smectic free-standing films are in the film plane, perfectly stacked. The local mesogen orientation is described by the c-director field, which points in the direction of the local tilt. Because $\vec c$ is a true vector, defects with half-integer topological strength are forbidden and the defects with the lowest elastic energy are those with topological strengths $S=+1$ and $-1$. The energy of each defect is proportional to the square of its topological charge \cite{Dafermos1970,Kleman2003}, thus defects with higher topological charges $|S_0| > 1$ have higher elastic energies than the sum of $N=|S_0|$ defects of charges $S=+1$ or $-1$, resp., with the same total topological charge. As a consequence, defects with strength $S_0>1$ ($S_0=2,3,4,\dots$) will decompose into $N$ defects of topological charges $+1$, under conservation of the total charge. 

	\begin{figure}[ht!]
		\centering
	 	\includegraphics[width=0.6\columnwidth]{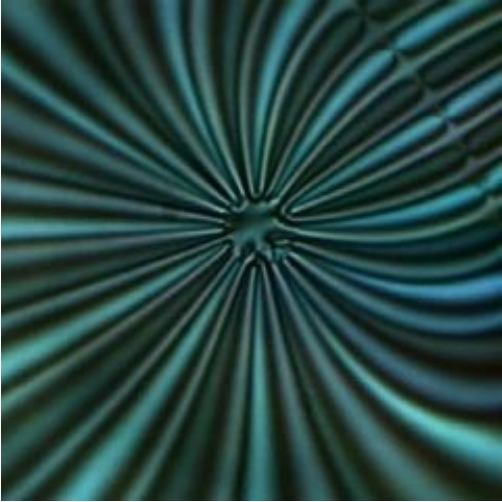}
		\caption{\label{fig:sunray} Example of the c-director texture around a
		hole in the smectic film (hole diameter approximately 7.5~$\mu$m) that contains 12 defects of topological strength $S=+1$ each. The defects decorate the border of the hole and they prevent the hole from shrinking to extinction.
		The image was taken under crossed polarizers, each individual defect creates four bright and four dark stripes in this sunray pattern.
		Image size 74.5 $\times 74.5~\mu$m$^2$.
		}
	\end{figure}

When $N$ defects of equal topological strength $+1$ are initially trapped within a narrow region, and then are released from this trap, they behave analogously to a decomposing single defect of topological charge $S_0$. Such an experiment was performed earlier, and an experimental technique to prepare a restricted region where the director field has a total topological charge $S_0>1$ was introduced \cite{Stannarius2016}. The technique basically exploits the fact that film regions with reduced number of smectic layers (so-called 'holes') can trap defects. The defect energy depends linearly upon the local film thickness, thus defects in thinner film regions are caged there. When one is able to produce a local area of reduced film thickness that contains a number of topological defects of the c-director field, this will often lead to a stable configuration where the line tension of the dislocations surrounding the hole is balanced by the defect repulsion. The mutually repelling defects cannot leave the hole into the thicker surrounding film regions, so they distribute themselves along the perimeter of the hole. This arrangement prevents the hole itself from shrinking and vanishing.
\begin{figure*}[htbp]
		\centering
	 	\includegraphics[width= 0.75\textwidth]{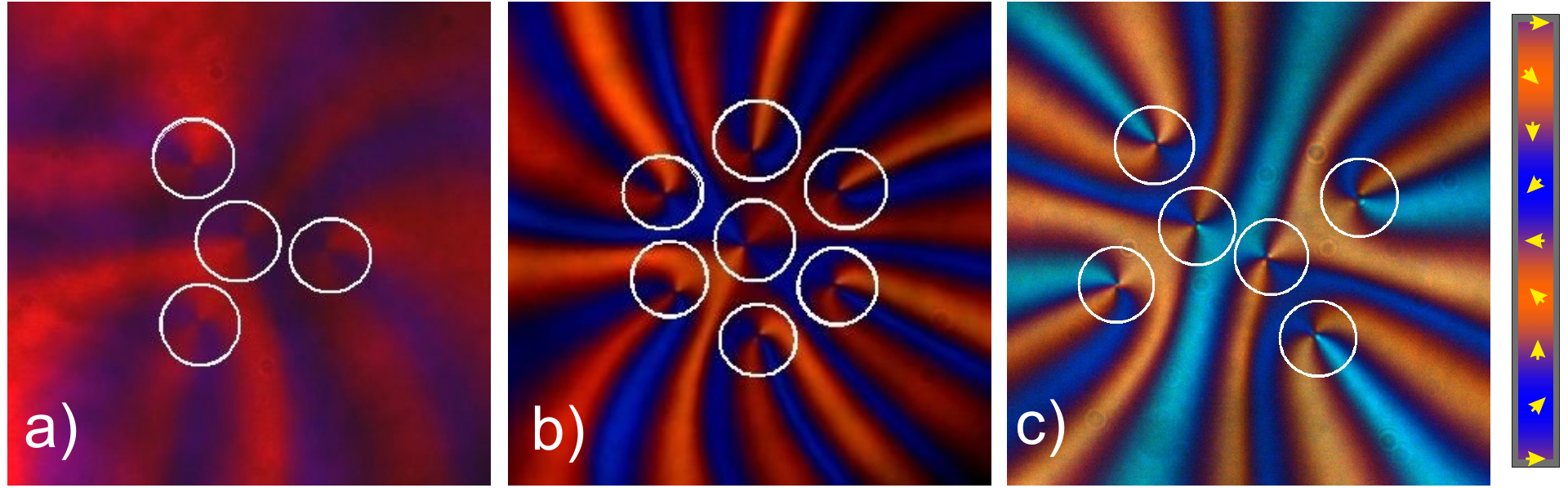}
		\caption{a-c) Examples of typical disintegration scenarios of defect clusters: a,b) One central defect at rest and $N-1$ nearly symmetrically departing defects for a) $S_0=4$ and b) $S_0=7$. c) shows a rare scenario of an $S_0=6$ cluster decomposing into four outer and two inner defects, all moving radially outward. The experimental images were taken with crossed polarizers and a $\lambda$ wave plate inserted diagonally. Blueish and reddish colors indicate regions where the c-director is along one of the diagonals, respectively. Defects are marked by white circles. The color bar to the right is a rough sketch of the relation between c-director orientation and film reflection color. Details of the texture colors depend on the film thickness, which increases from (a) to (c). 
		}\label{fig:def_examples}
	\end{figure*}

{
The material used in the experiment was a non-chiral room-temperature SmC  binary (50 vol\% : 50 vol\% ) mixture of
5-n-Octyl-2-[4-(n-octyloxy)phenyl]pyrimidine and
5-n-Octyl-2-[4-(n-Hexyloxy)phenyl]pyrimidine \cite{Stannarius2016}. All experiments were performed at room temperature.
Thin films were drawn across a frame with one moveable edge inside a THMS 400 hot stage. The width of the films was approximately 4~mm. The defect textures were observed using a Carl Zeiss AxioImagerPol polarizing microscope under crossed polarizers (horizontal and vertical in the images). In most of the experiments, a diagonal $\lambda=550$~nm phase plate was inserted to distinguish the two diagonal c-director orientations in the textures. The films around the trap had uniform thicknesses  roughly in the range from 300 nm to 3000 nm. 
After quenching the hole containing the defects, the film thicknesses were homogeneous in the observation area. 
}
 
Using this preparation method, we were able to create structures with total defect strengths $S_0$ between $+4$ and $+12$ (Fig.~\ref{fig:sunray})
in experiments, and there seems to be no principal limit to trap more defects \cite{Stannarius2016}. The technique to release the defects from their cage involves a quick in-plane compression of the film so that the hole is quenched and the trap is eliminated.

The unleashed identical defects repel each other, and the local cluster will thus spread rapidly. The initial state is a cluster of defects uniformly distributed on the periphery of a tiny circular region. Thus, one might naturally expect from symmetry considerations, that after the release of these defects, they all move away radially from the cluster center, uniformly distributed on a circle it. This scenario
was surprisingly never observed in our experiments. 
	
Instead, in the majority of experiments, the initial cluster decomposed with one defect remaining in the center and all others moving away radially [Figs. \ref{fig:def_examples}(a,b)]. In few cases, other scenarios were found like an $S_0=6$ cluster consisting of two inner and four outer, nearly symmetrically arranged defects that all moved radially away from the cluster center [Fig.~\ref{fig:def_examples}(c)]. Also, a cluster of two inner and six outer defects was found for $S_0=8$ as shown below. 
The reason for the preference of these structures, and the absence of others, will be given in the following.
We note that the absolute thickness of the homogeneous film itself after extinction of the hole does not influence the structure and dynamics of the decomposition process: The relevant quantities, e.g. the dissipation in the film and the elastic energies, both scale linearly with the film thickness.

The result of the previous experimental study \cite{Stannarius2016} was that the pinning of the c-director field in the vicinity of the core of a $+1$ defect is an essential feature that
determines the disintegration process.
In a strict one-constant elastic approximation, the director is not restricted to a tangential orientation near the defect core, it may have any offset angle, the defect can have a tangential or radial director field or any other phase. However, the elastic constant for bend
of the c-director field, $K_{\rm B}$ is actually considerably smaller than the constant for splay, $K_{\rm S}$. The experimentally determined ratio \cite{Stannarius2016} is $K_{\rm S}/K_{\rm B} \approx 2.2$. As a consequence, a defect with a radial
orientation of the c-director around its core would have a substantially larger elastic energy than a defect with purely tangential anchoring
of the c-director around the core. It would thus change to tangential orientation. The tangential defects represent the lowest energy configuration, and in fact, all $+1$ defects in the experiments had a tangentially pinned c-director (cf. Fig.~\ref{fig:def_examples}). Such defects can either have a clockwise or counterclockwise orientation of the c-director, with equal elastic energies. In ferroelectric smectic C$^\ast$ phases, this symmetry may be broken \cite{Bohley2007} and there may be a preference for one of the two winding senses, but not in the smectic C material used here. In Fig.~\ref{fig:def_examples}(a,b), the outer defects have the same winding sense, opposite to that of the central one.

\begin{figure}[htbp]
		\centering
	 	\includegraphics[width=0.6\columnwidth]{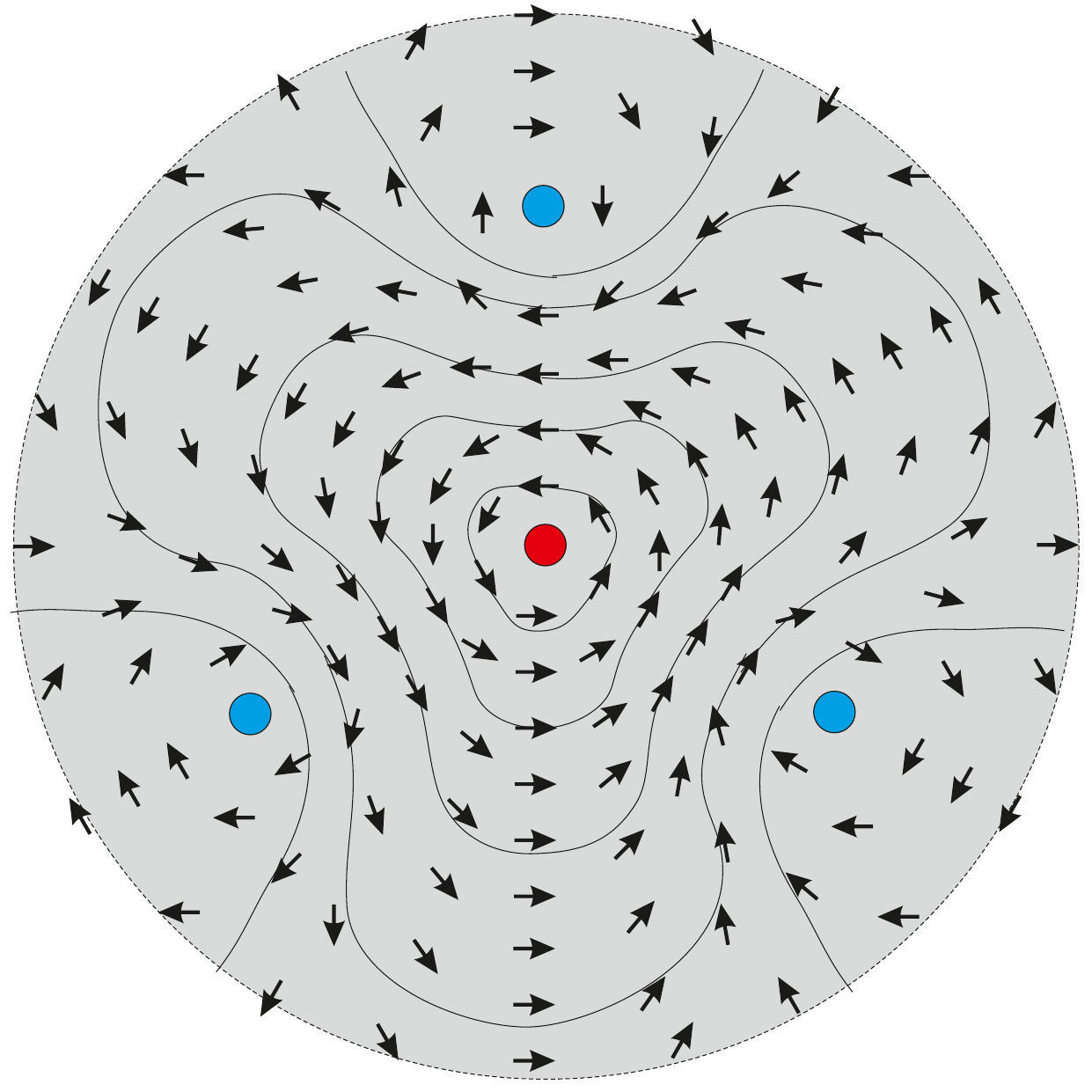}
		\caption{Sketch of the c-director field for the disintegrating $S_0=4$ cluster. The c-director is assumed counterclockwise around the central defect core and clockwise around the other three (color-coded in the sketch). The same configuration can also exist with the arrows reversed.
		Note the three inversion walls that necessarily form between the outer defects, indicated by the bent flow lines of the director field.
		}\label{fig:four}
\end{figure}
	
\section{Model for the defect configurations}

It was argued in Ref. \cite{Stannarius2016} that a construction of a cluster of three tangential $+1$ defects by simple superposition of the single defect solutions is possible only if they are arranged along a straight line. Otherwise, one has to introduce additional distortions of the director field to satisfy the tangential anchoring conditions near the defect cores. In contrast, when more than three defects are involved, there are arrangements of superimposed single-defect solutions that fulfill the anchoring conditions without the necessity to add further distortions. For example, a symmetric arrangement of four defects as observed experimentally [Fig.~\ref{fig:def_examples}(a)] can be constructed by a mere superposition of four individual $+1$ defects even if the defects are not lined up. 
Figure \ref{fig:four} sketches the c-director field for the case $S_0=4$, constructed as a linear superposition of the solutions for the director
deflection angles obtained from four individual defects. Superficially, it might appear as if the inversion walls between the outer defects have 
been additionally included, but they are actually the result of the superposition of the analytical single-defect solutions.
In absence of other director distortions, one can find simple analytical solutions for static defect configurations (i.e. assuming fixed defect positions) of this type in elastic one-constant approximation.
During the actual cluster disintegration, the central defect remains in the original position, the other three move radially outward.
Because of the spatial and temporal scaling characteristics of the involved dynamic equations for the  c-director and flow fields, the solutions are self-similar when all spatial dimensions are scaled with $t^{1/2}$, provided that the mutual defect distances are large compared to the defect core radii (This is in contrast to Ref.~\cite{Svensek2003} where the defects with overlapping cores were very close to each other and therefore different scaling factors were obtained.). This scaling behavior is fully confirmed in our experiments.

We suggest a model for the pattern selection based on the following simplifying assumptions:
\begin{enumerate}
\item The defect patterns evolve quasi-adiabatically, i.~e. they pass through states equivalent to static equilibrium solutions
of the director field with fixed positions of all defects. This assumption was also made for the description of annihilation dynamics \cite{Dafermos1970,Kleman2003}.
 \item The director fields for individual defects can be described in first approximation by the analytical one-constant solutions. We disregard
the fact that splayed regions are expanded and bent regions are actually quenched in our material. The only effect of elastic anisotropy is 
the tangential pinning of the director at the core.
\item After decomposition of the initial cluster, the disclinations are arranged in such a way that additional distortions are avoided, and they choose
patterns where the tangential pinning at each defect core is fulfilled by a mere superposition of single defect solutions.
\item Once these patterns have formed, they expand in a self-similar way. The scaling factor of all distances is proportional
to $t^{(1/2)}$, where $t$ is the time from the begin of the disintegration process.
\end{enumerate}

The close resemblance of the experimentally observed textures with those obtained from superimposed single-defect solutions will serve as the main argument to justify this model. 
As explained above, Figs.~\ref{fig:def_examples}(a) and \ref{fig:four} represent a structure that meets our conditions.
Other such solutions for higher $S_0$ are shown in Figs. \ref{fig:def_examples}(b,c), as will be demonstrated below.

In our 'quasi-one-constant' elastic model we assume that the director field far from the defect core can still be described satisfactorily by
the one-constant approximation, but the actually existing elastic anisotropy leads to the tangential pinning of the director at the cores of the $+1$ disclinations.
There are reasonable arguments for this: At the core, the elastic energy diverges and the difference between splay and bend plays a decisive role, while outside the cores, the elastic anisotropy will merely lead to certain local corrections but not to qualitatively new solutions.

\section{Disintegration dynamics}

The simplest model for defect interactions \cite{Dafermos1970} assumes that the elastic constants for splay and bend of the c-director are equal, flow is neglected, and the orientational deformations are assumed as a linear superpositions of two isolated
defects.
The force between two such defects separated by a distance $ r_{12} $ is \cite{Chandrasekhar1986,Kleman2003}:
\begin{equation}
f_{12} =  2 \pi K S_{1} S_{2} / r_{12}\label{eq:t1}.
\end{equation}
It acts along the separation vector $\vec{r}_{12}$. $K$ is a mean elastic constant, and
$S_{1,2}$ are the defect strengths.
Disclinations of the same sign, such as considered in our study, repel each other. This equation was derived for nematics but it can be used for the smectic C films as well with a proper redefinition of $K$.
A defect motion with velocity $V$ relative to the film material is counteracted by a drag force
$
f_{\rm drag} = 2 \gamma V \ln(3.6/Er)\label{eq:t2}$,
where ${Er} =\gamma \nu r_{\rm c}/K$ is the Ericksen number, $\gamma$ the rotational viscosity, $\nu$ a characteristic velocity scale, and $r_{\rm c}$ the defect core radius \cite{Chandrasekhar1986}. This adjusts the defect velocities in this overdamped system to
\begin{equation}
|V| = \frac{K}{\gamma \ln({3.6}/{Er}) r_{12}} =  \frac{D_{1}}{r_{12}}\label{eq:t3}
\end{equation}
In this simplified model without flow, the absolute defect velocity does not depend on the sign of $S$. It is inversely proportional to the separating distance $r_{12}$ (disregarding the slight velocity dependence in the logarithm).
For the simple case of $S_1=S_2=+1$, an integration of the velocity $ \dot r_{12}= 2v$ leads to
the square-root law
\begin{equation}
r_{12}(t) = 2\sqrt{D_{1}(t-t_{0})}\label{eq:4},
\end{equation}
where $t_0$ is the time when an initial $+2$ defect decomposes to generate the pair. The 'diffusion coefficient' $D_1$ is defined by Eq.~(\ref{eq:t3}). One may expand this simple model to groups of multiple defects assuming a linear combination of the individual forces. 
In the case of $N$ defects arranged with one central defect at rest and $N-1$ defects moving outward, the radial force on each outer defect would amount to $N f_{12}/2$ \cite{Stannarius2016,note2016}, with $r_{12}$ being its distance from the central defect.
It was shown that this is in contrast to the experimental findings \cite{Stannarius2016}. There may be several reasons, some of which we will discuss below.

It should be noted that Radzihovsky \cite{Radzihovsky2015} raised the objection that a moving defect may not necessarily have the same structure as a defect at rest relative to the film material. Experimentally, such effects have not been detected, presumably they are too small in the present experiments.

Finally, one also has to consider the effects of material flow on the defect dynamics:
It is well known that the motion of a disclination in the film plane causes material flow, in particular in the vicinity of defects with topological charge $+1$ \cite{Svensek2003}. This flow field has several consequences: First, it speeds up the motion of the $+1$ defect because the material locally flows in the same direction as the defect core is displaced relative to the film. Second, it leads to an asymmetry in the annihilation of defect pairs
\cite{Svensek2003}. 
In experiments, flow generated in the annihilation process of defect pairs
in freely suspended Smectic C films has been demonstrated using fluorescence microscopy and a photo-bleaching technique \cite{Missaoui2021}. In the present cluster disintegration scenario, however, an experimental confirmation of backflow has not been tackled yet.
Nevertheless, we hypothesize that near the positions of the outer $+1$ defects, the flow field is directed
outward, with the defect motion, thus increasing the defect speed. A compensating flow directed inward is expected in the inversion wall regions between the outer defects, in order to fulfill the continuity equation. When more defects are on the periphery, the distance between these defects 
becomes smaller and shear forces will dampen the supporting flow field, therefore it can be expected that the acceleration of the defects by flow
will be particularly effective for small $S_0$, whereas it will rapidly decrease with larger $S_0$.

\section{Geometrical analysis of groups of tangential +1  defects}
\label{sec:TheoryMultiple}
In the following geometrical considerations, we will neglect flow and consider the static solutions in one-constant approximation.
The advantage of the one-constant model without flow is that it yields a crude but analytical approximation for the patterns and their
dynamics. The square-root law, Eq.~(\ref{eq:4}), will remain effective even when the neglected terms related to material flow and elastic anisotropy  are included.

The experiments show that the geometry of the growing defect pattern, once it has formed, remains constant. Thus, it is sufficient for the following discussion to consider an arbitrarily chosen snapshot of the  
video sequences and discuss the defect arrangement, in order to obtain the full description of the spatio-temporal behavior. 

We will discuss, which types of defect configurations can be constructed under the model assumptions given above, as superpositions of single-defect equilibrium solutions.
For the analysis of these configurations, a 2D vector field $\vec n_0(x,y)=[n_x(x,y),n_y(x,y)]$ is considered in the $(x,y)$ plane,
and the azimuthal angle $\theta$ is defined such that
$n_x=|\vec n_0|\cos\theta(x,y)$, $n_y=|\vec n_0|\sin\theta(x,y)$.
$N$ defects of identical topological charges $S_i=+1$ ($i=1,2,3,\dots N)$ are distributed in the plane. 
In the vicinity of the core of
the $i-${th} defect, $\theta = \varphi_i+\theta_i$, where $\varphi_i$ is the azimuthal angle of the spatial vector from the defect center and $\theta_i$ is a phase angle.
Tangential orientation means that $\theta_i = (n_i+1/2)\pi$,
(integer $n_i$). For odd $n_i$, the rotation of the director around the core is counterclockwise, and for even
$n_i$ this rotation is clockwise. In the system considered here, 
all defects have a tangential orientation of the director near the core as a result of the minimization of the free energy when bend
is favored over splay. In a material where splay deformations are energetically favored, an analogous treatment is possible. The solutions are the same except that an offset of $\pi/2$ has to be added to $\theta(x,y)$.

\begin{figure*}[htbp]
a) \includegraphics[width=0.44\columnwidth]{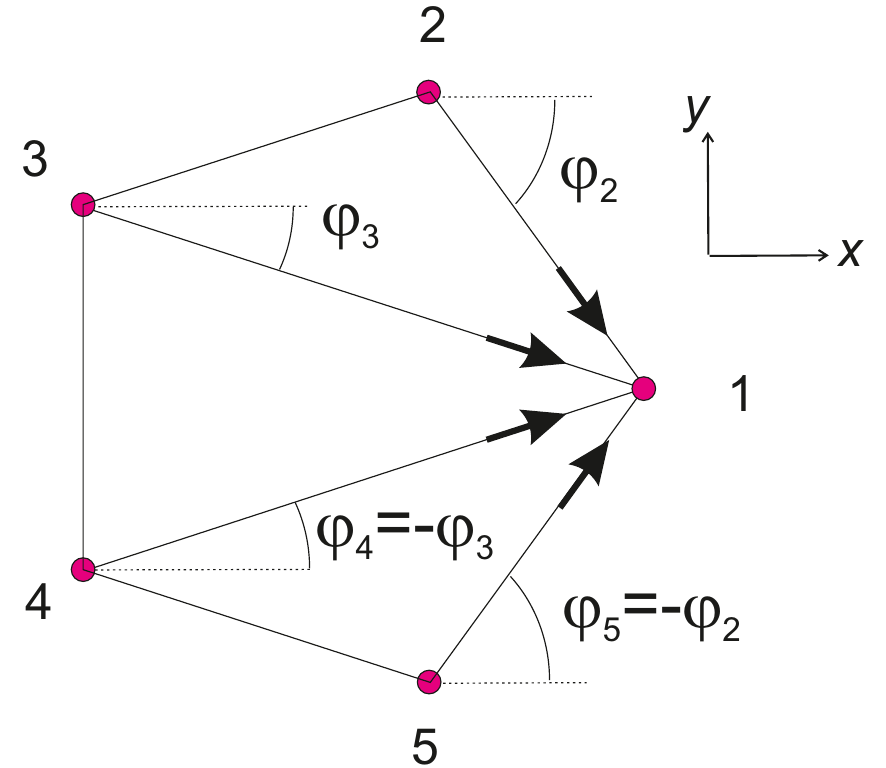}\hfill
b) \includegraphics[width=0.44\columnwidth]{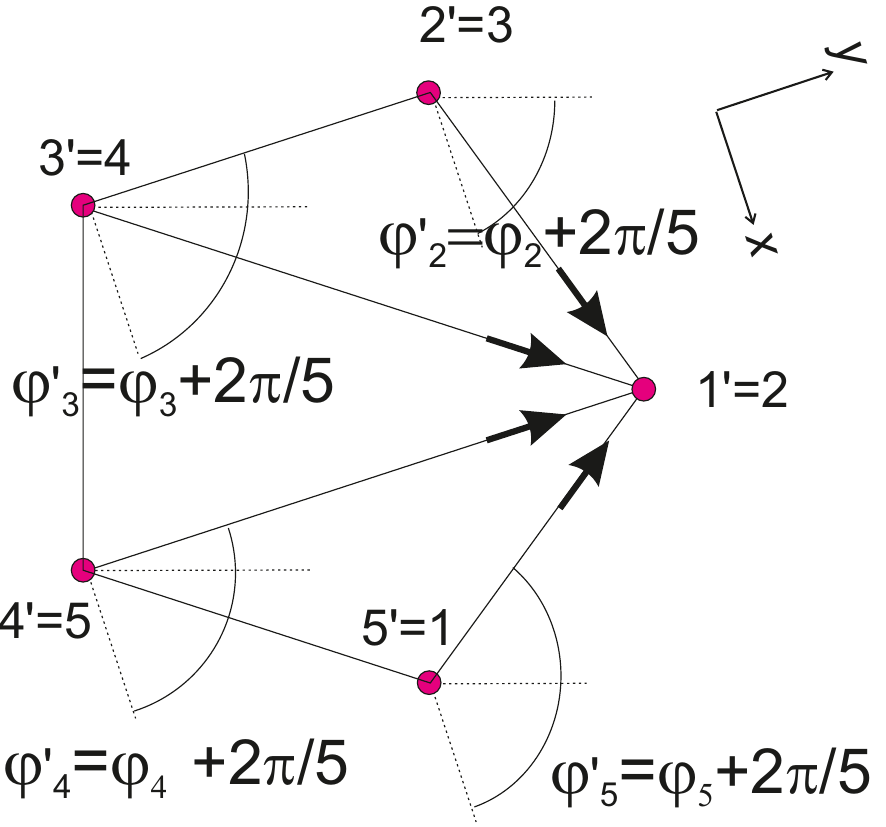}\hfill
c) \includegraphics[width=0.44\columnwidth]{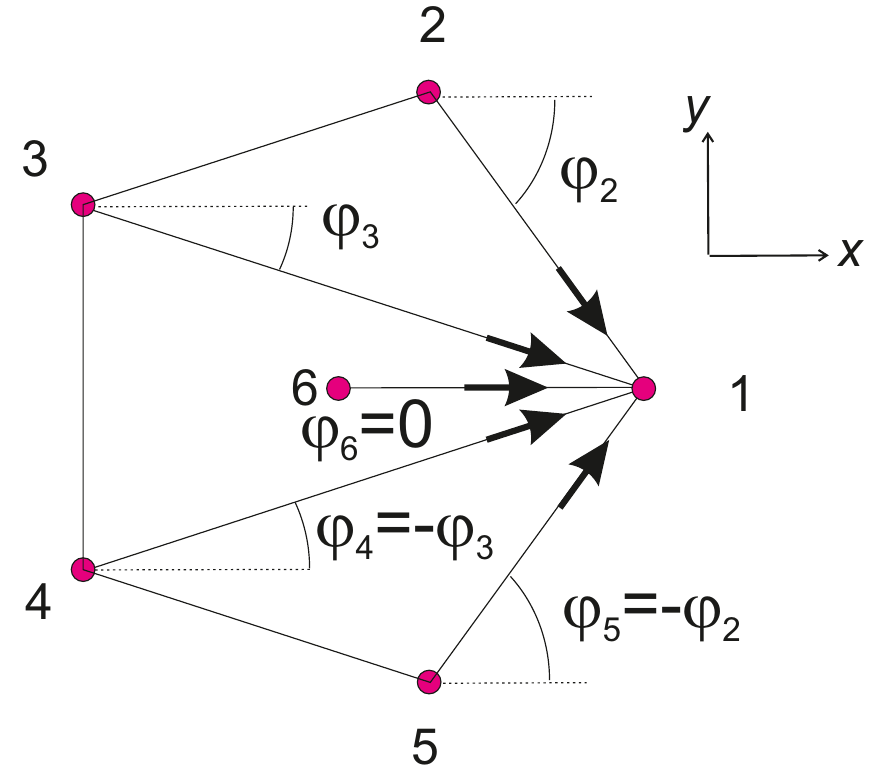}\hfill
d) \includegraphics[width=0.44\columnwidth]{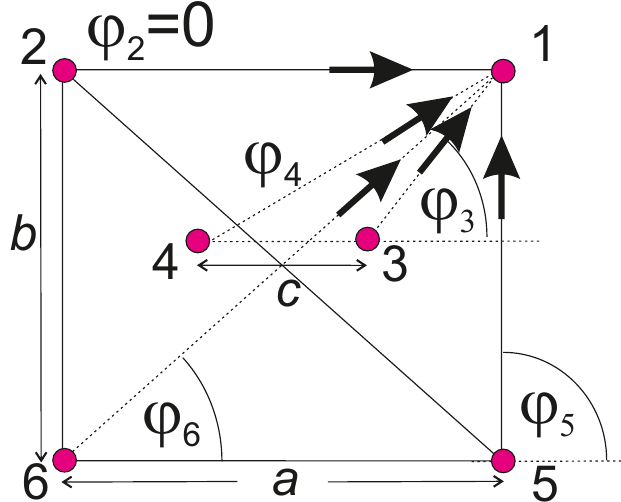}
	\caption{Sketches of symmetric arrangements of $+1$ defects.
a) five outer defects, no central defect. The contributions of the defects to the phase $\theta_1$ are sketched by arrows.
b) same for the defect at position 2
c) central defect added
d) six defects, two of them in central position, four outer defects. The contributions to $\theta_1$ are indicated by arrows.}
\label{fig:symmetry}
\end{figure*}

In absence of other director distortions, the c-director angle $\theta(x,y)$ in the film plane is additively composed of the sum of $N$ contributions $\varphi_i$ and a constant offset
 $\theta_{\rm o}$. The phase
angle $\theta_k$ of a given defect $k$ thus results from the contributions of the other $N-1$ defects and the offset:
\begin{equation}
\theta_k=\theta_{\rm o}+\sum_{i\neq k} \varphi_i .
\label{eq:1}
\end{equation}
The offset $\theta_{\rm o}$ can be used to adjust the tangential anchoring for all defects.
We can therefore limit the analysis
to the search for configurations where all $\theta_i$ are equal modulo $\pi$,
and reserve the offset to make them equal to $(n_i+1/2)\pi$

In general, the superimposed director fields of $N$ defects at arbitrary positions will not meet this condition. Yet there are exceptions, and we will demonstrate a few of them next. Since the experimental patterns are highly symmetric, we will consider only symmetric cases here.
Three $+1$ defects in a line are such a special solution. The rotation sense of the central defect is opposite to that of the outer two ones. In any other
constellation of the three defects, the phases cannot be matched to be equal modulo $\pi$.

Let us now analyze configurations without central defect, where the outer
$N$ defects are arranged symmetrically in a distance $R$ from the center in angular directions $2\pi (i-1)/N $
(Fig.~\ref{fig:symmetry}a,b).

In that case,
a superposition of the distortions yields {\em different} phases for all five defects according to Eq.~(\ref{eq:1}). This can be derived straightforwardly. We demonstrate it for the
example $N=5$, shown in Fig.~\ref{fig:symmetry}a:
The sum of the four contributions to defect 1 is $\theta_1=\varphi_2+\varphi_5+\varphi_3+\varphi_4= \theta_{\rm o}$, because symmetry requires $\varphi_4=-\varphi_3$ and $\varphi_5=-\varphi_2$
when defect 1 is located in $x$ direction from the center. For the second defect, $k=2$, it is easiest to find the phase by rotating the coordinate system by $2\pi/N$ and re-labeling the defects as shown in Fig.~\ref{fig:symmetry}b. Now, each of the four (i. e. $N-1$) azimuth angles $\varphi_i, (i\neq k)$ is increased by the rotation angle $2\pi/N$, yielding the total phase $2\pi (N-1)/N + \theta_0$ for defect 2. In the same way, one finds that all defects $k=1,2,3,4,5$ obtain the phases
$$\theta_k =2\pi (k-1)(N-1)/N + \theta_0.$$ Thus, all five phases $\theta_k$ differ by values that are not integer multiples of $\pi$, and an all-tangential alignment is impossible. This applies for arbitrary $N>2$.

One can nevertheless force configurations as solutions of the director field where all $N$ defects in such an arrangement become tangential, but this would require the introduction of additional deformations of the director field, and thus additional elastic energy.
It should be noted that in the experiments \cite{Stannarius2016}, we have never found this type of configurations after the
disintegration of the central defect cluster. This is a strong indication that the system tries to avoid the additional elastic distortions. 
It applies even though the system is naturally not in equilibrium when the defect cluster spreads.

When one analyzes the second case, with a symmetric arrangement of $N-1 $ defects outside and one central defect, the situation changes completely.
For $N=3$, these are the trivial three lined-up defects.
The case $N=6$ is shown in Fig. \ref{fig:symmetry}c. The outer defects $2,3,4$ and $5$ contribute a total phase of zero to $\theta_1$, as indicated in the figure. On the other hand, these defects
$k=2,3,4,5$, as in the previous case, gain the contributions $2\pi (k-1)(N-1)/N$ to $\theta_k$ from the other outer defects. However, the central defect additionally contributes a phase $2\pi (k-1)/N$ to each of them so that the outer
defects obtain the phases $\theta_k = 2\pi (k-1)+\theta_{\rm o}$. Thus, the $\theta_k$ are integer multiples of $2\pi$ plus the offset. For the central defect, the sum of all
outer defects yields $\pi$, because contributions of defects $2\dots N-1$ cancel each other for symmetry reasons. Only the contribution of the first
defect, $\varphi_1=\pi$, remains besides $\theta_o$. As a consequence, the arrangement shown in Fig.~\ref{fig:symmetry} and all structurally similar arrangements
for $N\ge 3$ can be constructed by the superposition of the individual single-defect solutions without adding other deformations. With a proper choice of
$\theta_{\rm o}$ (depending on the orientation of the coordinate system) one can make all defects tangential.

Another configuration that was also observed in the experiment [Fig.~\ref{fig:def_examples}(c)] is that of two inner defects and a group of outer defects, shown exemplarily for
six defects in Fig. \ref{fig:symmetry}d. It is easy to see that defects 3 and 4 have phases
$\theta_3=\theta_{\rm o}$ and $\theta_4=\pi+\theta_{\rm o}$ (the contributions of the four outer defects to the phase angles cancel each other pairwise). The phases of the two inner defects thus differ by $\pi$, so when one of them is tangential clockwise, the other one is anticlockwise. The phase $\theta_1$
is composed of the contributions $\theta_0$, $\varphi_5=\pi/2$, $\varphi_2=0$ and the sum
$\varphi_3+\varphi_4+\varphi_6$. The latter three angles can be expressed by the ratios of the sides $a,b,c$ as
$\varphi_6=\arctan{(b/a)}$, $\varphi_3=\arctan{[b/(a-c)]}$, and $\varphi_4=\arctan{ [b/(a+c)]}$ (see Fig. \ref{fig:symmetry}d). In order to bring the phase $\theta_1$ to
 some $n\pi +\theta_{\rm o}$ with an integer $n$, one needs to solve the transcendental equation
\begin{equation}
\frac{\pi}{2} + n\pi= \arctan \frac{b}{a} + \arctan \frac{b}{a-c} +\arctan \frac{b}{a+c} .
\label{eq:atan}
\end{equation}

This equation can be simplified using the addition theorem for the $\arctan$ function, $$\arctan X + \arctan Y = \arctan [(X+Y)/(1-XY)]$$
to
\begin{equation}
\arctan\frac{-b^3/a^3-c^2b/a^3+3b/a}{1-c^2/a^2-3b^2/a^2}=\frac{\pi}{2} + n\pi.
\label{eq:atan2}
\end{equation}
Since the $\arctan$ yields $\frac{\pi}{2} + n\pi$ for the argument $\pm \infty$, the denominator in the argument must vanish.
The solution is obviously
 
\begin{equation}
c = \sqrt{a^2-3b^2}.
\label{eq:c}
\end{equation}

\begin{figure}[htbp]
a) \includegraphics[width=0.38\columnwidth]{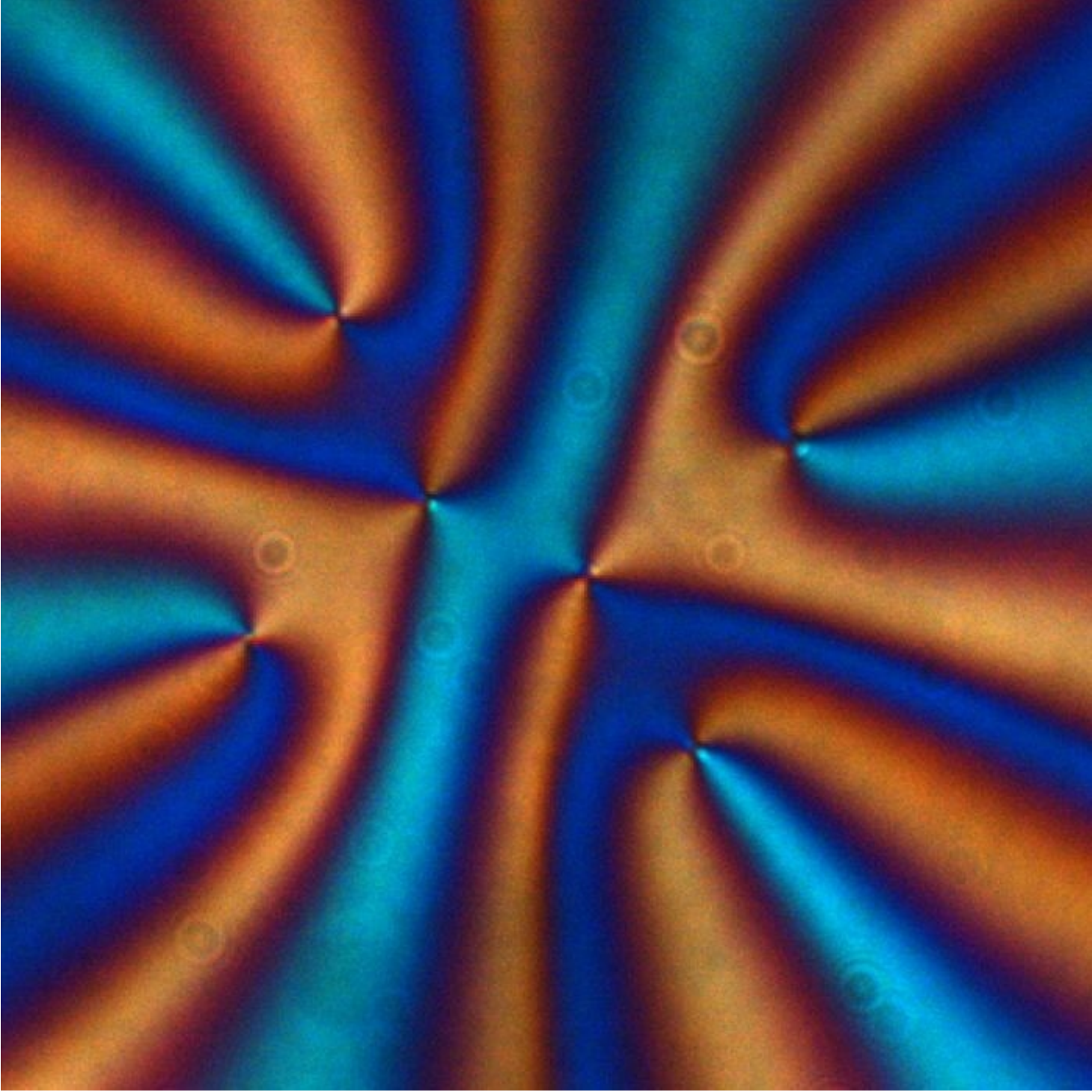}
b) \includegraphics[width=0.38\columnwidth]{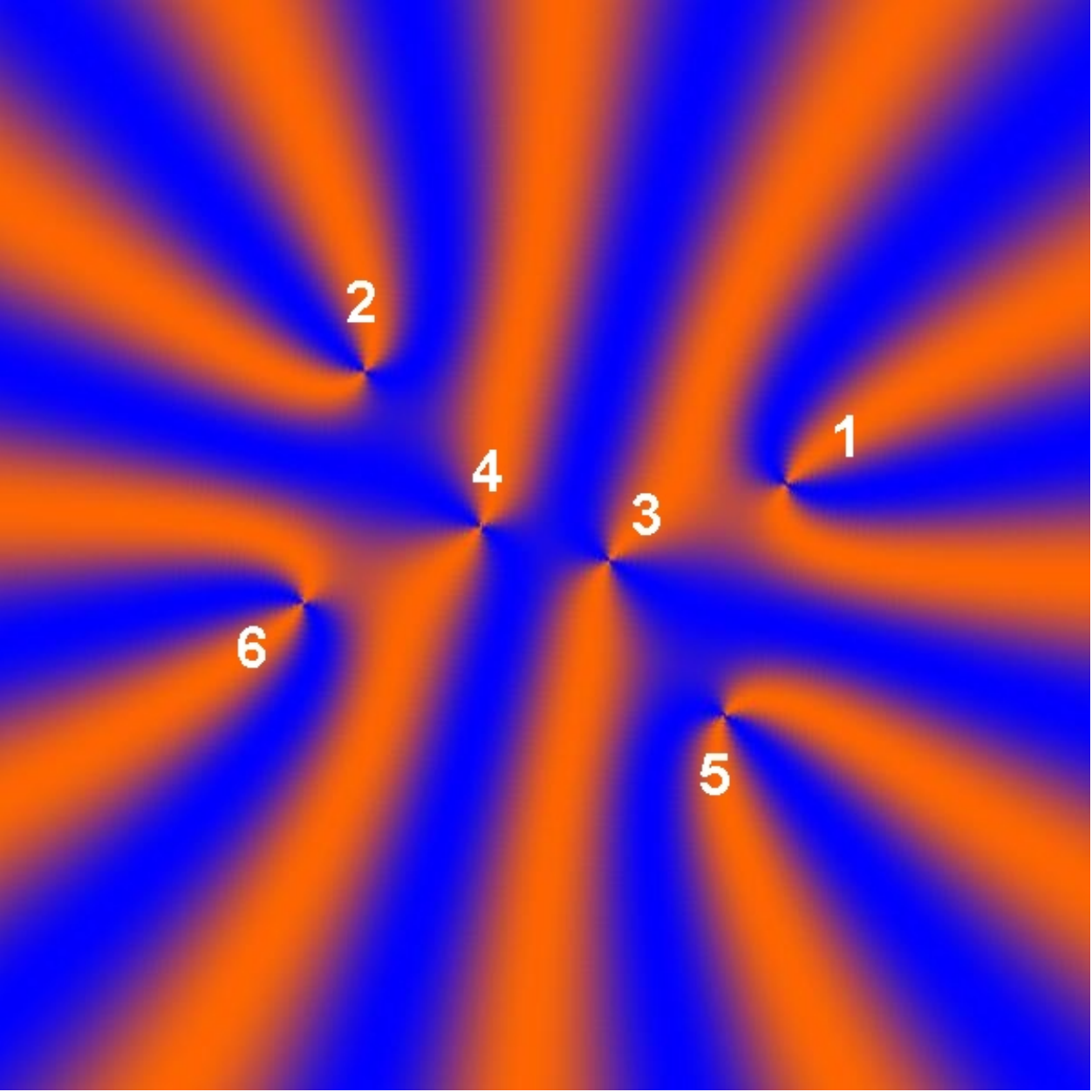}
\hfill\includegraphics[width=0.11\columnwidth]{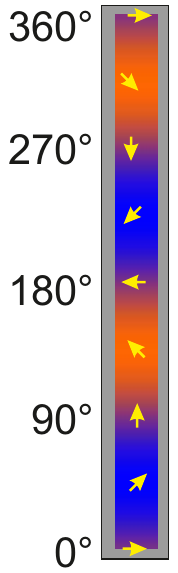}
	\caption{Comparison of a 2 + 4 cluster in experiment and simulation. a) is the experimental texture. b) was simulated to roughly
represent the model texture. The outer defects were placed in positions $(\pm a/2,\pm b/2)$ as in the experiment, and the two inner defects
at $(\pm c/2,0)$ with $c$ computed from Eq. (\ref{eq:c}). The colors were chosen as a crude approximation to the optics under crossed polarizers and
a diagonal phase plate (see text).}
\label{fig:six}
\end{figure}

Using simple symmetry arguments, one can find the same condition for the adjustment of phases $\theta_2$, $\theta_5$ and $\theta_6$.
When the distances $a,b,c$ between the defects fulfill this equation, then one can achieve tangential anchoring for the six defects by a simple
superposition of single defect solutions plus offset. Note that for $b\rightarrow 0$, the defects 1,3, and 5 come very close
to each other and the elastic energy of such a configuration will naturally become very large. For the other limit, {$b\rightarrow a/\sqrt{3}$},
$c$ goes to zero and the two inner defects approach each other very closely, with the same effect. The system obviously chooses a
configuration of $c/b$ where the total potential energy is near an optimum. This is the case in the experiment. Figure
\ref{fig:six} shows the experimental observation again (a) and the simulated texture (b) for the experimentally determined ratio $a/b\approx 1.83$ and $c$ determined from
Eq.~(\ref{eq:c}), $c/b\approx 0.6$. Instead of the director field, we show a color presentation where the intensity of the red channel of the image was set
proportional to $1+\sin\theta$, the blue channel to $1-\sin\theta$, the green channel to $0.4 (1+\sin\theta)$. The color scale at the right of Fig.~\ref{fig:six} visualizes the relation between the director and the simulated  texture color. This reproduces the positions of the diagonal (blueish and reddish arms in the experimental textures. 
Evidently, not only the defect positions but also the distortions that were generated by linear superposition of the six individual solutions are in satisfactory agreement. One has to keep in mind that we made two simplifying assumptions here: first, we assumed that the director field for a given defect arrangement is an equilibrium solution, and second, we
used the solutions for one-constant approximation which we know is not exact here because $K_{\rm S} > K_{\rm B}$.
The sense of rotation of the c-director is the same for defects 1, 4 and 5 in Fig.~\ref{fig:symmetry}d, and opposite for the other three defects 2, 3 and 6.

The arrangement for six defects shown in Fig. \ref{fig:six} can thus be explained easily by purely geometrical considerations, except for the experimentally selected ratio $a/b$ which is presumably related to a local minimum of the elastic energy.

\begin{figure}[htbp]
a) \includegraphics[width=0.38\columnwidth]{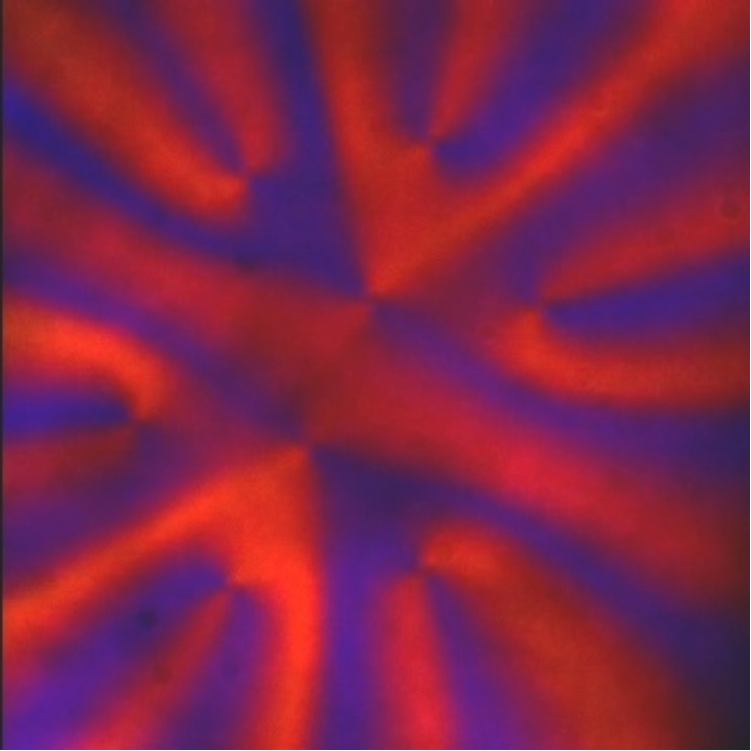}
b) \includegraphics[width=0.38\columnwidth]{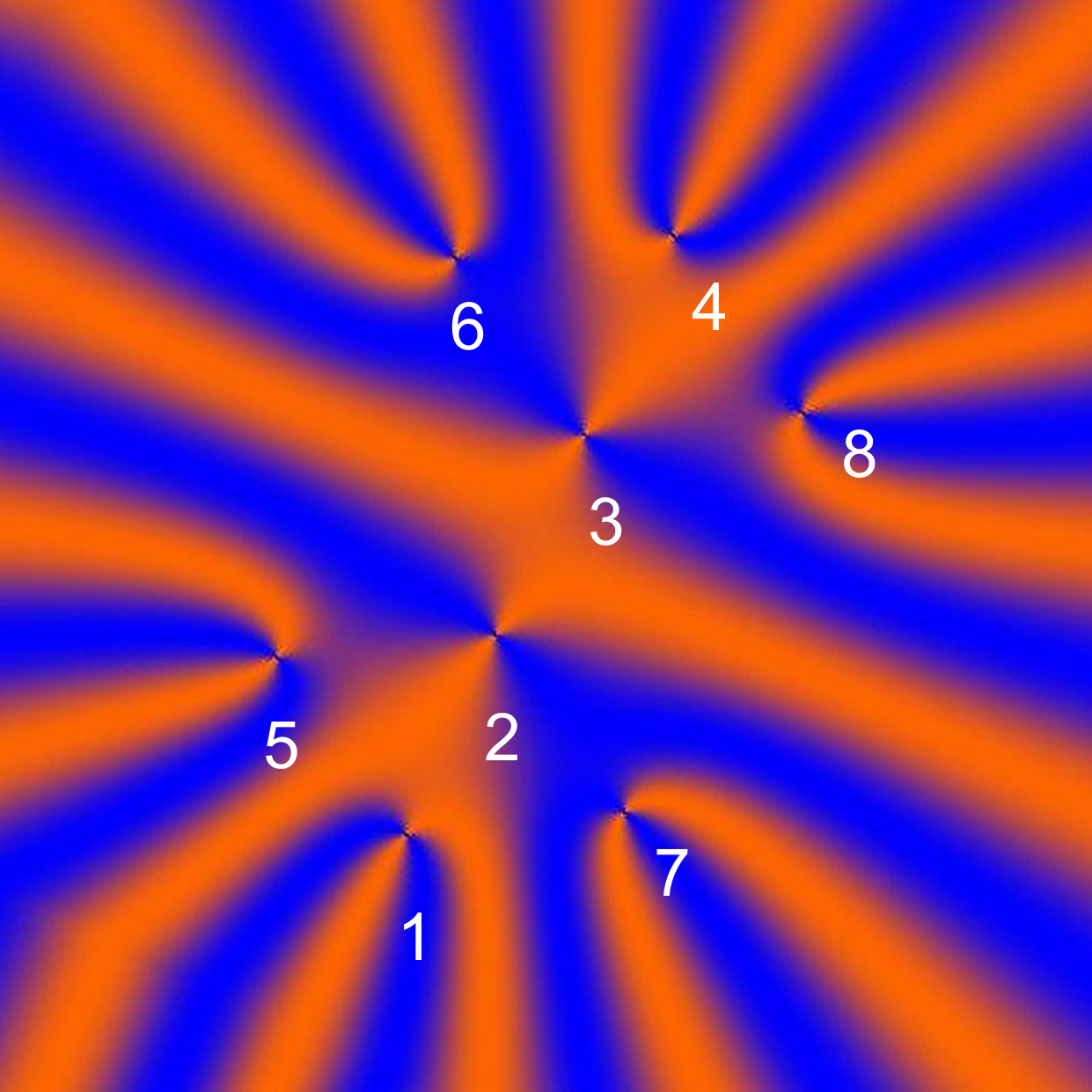}
\hfill\includegraphics[width=0.11\columnwidth]{cbar.pdf}
	\caption{Comparison of a 2 - 4 - 2 cluster of eight defects in experiment and simulation. a) is the experimental texture. b) was simulated to roughly represent the texture. Four defects were distributed equidistantly on a straight line. The other four defects were placed such that they
formed equilateral triangles with the first two and with the last two defects on the straight line, respectively. All direct neighbors thus have the same
mutual distances (sketched by dashed white lines).
Colors as in Fig.~\ref{fig:six}.}
\label{fig:eight}
\end{figure}

Another structure that also lacks a central defect but fulfills the requirements to be a mere superposition of single defect solutions was found coincidentally in the experiment for $S_0=8$. It is shown
in Fig.~\ref{fig:eight}. The experimental pattern suggested that one should seek for a superposition with four defects
$1,2,3,4$ distributed
equidistantly on a straight line (distance $\ell$), and an appropriate addition of defect pairs $5,6$ and $7,8$ on each side of the central line.
We found that defects $5$ and $6$ have to have a separation $2\ell$ and form two equilateral triangles $1,2,5$ and $3,4,6$ with side length $\ell$, respectively.
At the opposite side of the central line, $7,1,2$ and $8,3,4$ in a similar arrangement form another two equilateral triangles.
The comparison with the experiment shows that the two configurations are nearly identical. Small deviations may occur as a consequence of the
elastic anisotropy which was disregarded in the simulated pattern. One can calculate the phases $\theta_i$ ($i=1,2,\dots 8$) using the positions
described above in a one-constant model to confirm the correctness. We did not search for further geometrical solutions but we hypothesize that 
other special configurations may exist, particularly for larger $S_0$, which also meet the requirements listed above.

In all cases described above, the experimentally observed dynamic defect patters simply expanded self-similarly with a scaling factor proportional to the square-root of time,
until the defects approached the film boundaries or layer steps or other disclinations that might coincidentally be present in the film. Thus, the
snapshots shown are representative for the complete cluster disintegration process. 
The square-root law evidences that the elastic energies decrease inversely proportional to the length scale $\ell$ of the pattern.

\section{Discussion and Summary}
\label{sec:discsum}

It has been demonstrated that the textures observed during the disintegration of defect clusters with topological charges $S_0 \ge 4$ can be described
satisfactorily by a sequence of quasi-static equilibrium configurations that are passed in an adiabatic way. Even though the elastic anisotropy 
was disregarded in our analytical model, and used only for the tangential director pinning near the cores, the resemblance of analytically derived and experimentally observed textures is convincing. The total defect strength $S_0$ of the cluster determines the geometry of the created pattern that expands continuously in a congruent, self-similar fashion.
The c-director field can be described in reasonably good approximation by the superposition of the individual point defects without addition of surplus director deflections. The particular geometrical arrangements of the defects are enforced by the condition that the director is pinned tangentially near the defect cores. In a one-constant approximation, the configurations as well as their textures are obtained analytically.
When the two elastic constants for splay and bend differ, then details of the texture differ but the global structure is the same. Apart from the forced pinning of the director near the defect cores, the actual elastic anisotropy has no qualitative consequences for the observed patterns. This is evident, e.~g., in the satisfactory agreement between the experimental and calculated images in Figs. \ref{fig:six} and \ref{fig:eight}.

In the experiments, almost all disintegrating clusters formed patterns with a single defect remaining at the original cluster position and
$N-1$ symmetrically departing defects around. The winding sense of all peripheral defects is the same, opposite to that of the central one.

Defect configurations where all single defects were distributed uniformly on a circle without a central defect cannot be constructed under the condition of tangential
pinning, and they were never observed in our experiments. 

In some cases, the disintegration led to two inner and $N-2$ outer
defects in the experiment. The two inner defects have opposite winding sense and the director field has mirror symmetry with respect to a line perpendicular to the connection of the two inner defects. An example was shown in Fig.~\ref{fig:six}.
An initial $S_0=6$ cluster could also decay
theoretically into one central and five peripheral defects according to our model, even though we could not obtain experimental evidence for this type.

In Fig.~\ref{fig:eight}, a similar structure was shown for $S_0=8$. In this figure, defects with odd/even numbers have the same winding sense, respectively. This special arrangement also fulfills our conditions for the superposition model, provided that the required
geometric ratios are obeyed. Note that, because of the symmetry properties of the latter two patterns without a single central defect, these solutions exist only for even values of $S_0$. 

In the experiments, we did not find any other arrangements than the two principal types with one or two inner defects, and thus we have not searched for other analytical defect configurations that might meet our
conditions. Yet it cannot be excluded that such solutions exist, in particular when more than 12 defects are involved. Since all experiments started with a cluster where the individual defects were arranged at the periphery of a small circular hole, all patterns break the initial symmetry of the cluster. The reason for that must be sought in the 
avoidance of energetically costly additional director distortions.

The present model is by far not complete.
In a more realistic dynamic model, one has to consider the induction of flow fields by the moving defects, and also the anisotropy of the elastic constants.
This task was not tackled here, and is presumably not achievable analytically.
It will not lead to qualitative changes of the model.
One has to start with the known continuum equations for the c-director field. In the 2D film geometry, these can be directly derived from
the nematic Leslie-Ericksen equations. Then, one can adopt the ideas of Sven\v{s}ek and \v{Z}umer~\cite{Svensek2003a} for that purpose, and use the projection $\vec{n_0}=(n_x,n_y)=(\sin\beta\cos\theta,\sin\beta\sin\theta)$ of the director
$\vec{n}$ onto the $xy$-layer plane instead of the c-director. This vector is in the same direction as $\vec c$ but has a length that varies with the sine of the local smectic tilt angle $\beta$. The latter serves as an order parameter variable. This tilt angle and thereby the
length of $\vec{n}_0$) can deviate locally from its equilibrium value in the vicinity of the defect cores.
This task is in progress \cite{Harth202x}.

\section*{Conflicts of interest}
There are no conflicts to declare.

\section*{Acknowledgements}
 	The German science foundation (DFG) is acknowledged for support within projects STA 425/42-1 {and HA 8467/2,} and the German Space Administration (DLR) is acknowledged for support within grant 50WM2048.



\balance



\providecommand*{\mcitethebibliography}{\thebibliography}
\csname @ifundefined\endcsname{endmcitethebibliography}
{\let\endmcitethebibliography\endthebibliography}{}

\end{document}